\documentclass[letterpaper,12pt,prl,onecolumn,aps,superscriptaddress]{revtex4-1}
\usepackage{fullpage}
\usepackage{amsmath}
\usepackage{amsfonts,wasysym}
\usepackage{amssymb}
\usepackage{graphicx}
\usepackage{color}
\usepackage{slashbox}
\usepackage{longtable}
\usepackage{array}
\usepackage{dashrule}

\newcommand{\Vp}{{V_{P}}}
\newcommand{\Vf}{{V_{F}}}
\newcommand{\Vd}{{V_{D}}}

\begin{document}
(draft, \today)
\title{\Large{Hertz beyond expectations}}
\author{Andong He}
\affiliation{Nordic Institute for Theoretical Physics, 10691 Stockholm, Sweden}
\affiliation{Yale University, New Haven, Connecticut, 06520-8109, USA}
\author{John Wettlaufer}
\affiliation{Nordic Institute for Theoretical Physics, 10691 Stockholm, Sweden}
\affiliation{Yale University, New Haven, Connecticut, 06520-8109, USA}
\affiliation{Mathematical Institute, University of Oxford, Oxford OX1 3LB, UK}

\begin{abstract}
We examine the validity of Hertz's linear elastic theory for central collisions using a viscoelastic model. This model explains why Hertz's theory is accurate in predicting the collision time and maximum contact area even when 40\% of the kinetic energy is lost due to viscous dissipation. The main reason is that both the collision time and maximum contact area have a very weak dependence on the impact velocity. Moreover, we show that colliding objects exhibit an apparent size dependent yield strength, which results from larger objects dissipating less energy at a given impact velocity.
\end{abstract}

\maketitle

\section{Introduction}


Collisions are ubiquitous in nature.  The theory of linear elasticity developed by Heinrich Hertz has been widely used to describe the collisional dynamics between two elastic bodies (see e.g., \cite{Chokshi93,Gugan00,Hertz1882,Higa98,Johnson85,Leroy84,Tabor48}). Because of the linear elastic nature of Hertz's theory, the compression and recoil dynamics are symmetric, and the coefficient of restitution (COR), defined as the ratio of recoil velocity to impact velocity, is unity. Hence, there is no energy loss. 


Nevertheless, inelastic collisions are encountered much more often in nature, including sports \cite{Cross99}, geological saltation \cite{Carneiro13}, dynamics of granular gases \cite{Schwager98}, and growth of planetesimals \cite{Armitage10,Wettlaufer10}. In these processes, the loss of initial kinetic energy can occur through three principal channels: (1) vibrational modes of surface waves,  (2) plastic deformation or fracturing, and  (3) viscous dissipation due to internal friction \cite{King11,Kuwabara87}. When the impact velocity $V$ is much smaller than the speed of sound in the material, the energy conversion to surface waves is negligibly small \cite{Chokshi93, Mclaskey10,Reed85}. Similarly, plastic deformation only occurs for large $V$, when the maximum compressive stress exceeds the yield strength of the colliding bodies \cite{Tabor48,Thornton97}. Therefore, viscous dissipation is the only dominant means of energy loss for collisions at relatively low velocity.


Due to its elegance and compactness, it is tempting to apply Hertz's theory to weakly inelastic collisions. Indeed, the theory is found to be accurate in predicting the collision time and maximum contact area even when 40\% of the initial energy is lost \cite{Gugan00,Kennelly11,Lifshitz64,Villaggio96}. Gugan conjectured that this unexpected result is due to the fact that the compression stage is almost elastic and energy loss occurs mainly during recoil of the colliding objects \cite{Gugan00}. He rationalized that since the maximum contact area and the time taken to reach it are determined before the recoil takes place, they are not strongly influenced by how much energy is lost afterwards.


Here we use a viscoelastic model to study the energy loss due to viscous dissipation during a central collision. We find, in contrast to Gugan's conjecture, that the energy loss before and after the maximum compression are comparable. Moreover, the typical measurable collision quantities (i.e., collision time, maximum contact area, and maximum compression) are substantially less sensitive to the energy loss than the COR is. Therefore,  they do not deviate appreciably from Hertz's predictions even when the COR is much less than unity. Such deviations can be expressed as asymptotic series in a dimensionless number which characterizes the ratio of the viscous energy dissipation to the total energy.


We also analyze the commonly observed phenomenon that larger colliding objects fracture at a lower impact velocity, which is often taken as an indication that they are weaker than smaller objects. This can not be explained by either Hertz's theory which predicts a size independent maximum pressure \cite{Johnson85}, or an elastic-perfectly plastic model \cite{Thornton97}. Two alternative mechanisms have been proposed in the literature to explain this size effect: (1) Larger objects typically contain a larger number of defects and are thus more vulnerable in high-energy collisions (\emph{defect weakening}), 
and (2) Materials undergoing a larger strain-rate deformation tend to have a higher strength (\emph{strain-rate hardening}) \cite{Jones07}. We hypothesize that, when viscous energy dissipation is considered, the size dependence of the yield strength need not be defect weakening or strain-rate hardening. During compression the viscosity is always present to dissipate energy, so when the maximum pressure is achieved, its value is lower than that given by Hertz's theory. Since the fraction of energy loss depends on the size of the colliding objects, so does the actual maximum pressure. We test this hypothesis using the viscoelastic model and find good agreement with experimental data. 


\section{The Hertz theory}


Consider two isotropic and homogeneous spheres undergoing a central collision. For each sphere $i,(i=1,2)$, let $M_i$, $R_i$, $Y_i$, $\nu_i$ be its mass, radius, Young's modulus, and Poisson's ratio, respectively. From Hertz's elastic theory the collision time $T_e$ and maximum contact area $A_e\equiv\pi a_e^2$ ($a_e$ is the maximum contact radius) are given as
\begin{subequations}\label{basqua1}
\begin{equation}\label{elatime1}
	T_e=2.8683\left(\dfrac{M^2}{RVE^2}\right)^{1/5},
\end{equation}
\begin{equation}\label{elarad1}
	A_e=\pi\left(\dfrac{15MV^2R^2}{16E}\right)^{2/5},
\end{equation}
\end{subequations}
where $M=M_1M_2/(M_1+M_2)$, $R=R_1R_2/(R_1+R_2)$, and $E=1/\left[(1-\nu_1^2)/Y_1+(1-\nu_2^2)/Y_2\right]$ are the effective mass, radius, and Young's modulus, respectively. The case of a sphere-plane collision can be recovered by letting $M_2$ and $R_2$ become infinitely large. The maximum compression is given by the geometric relation $\delta_e= a_e^2/R$, and the maximum contact force $F_e$ and pressure $P_e$ can also be expressed as functions of $V$ and the material properties \cite{Hertz1882,Johnson85}.


\section{A viscoelastic model}


A modification of Hertz's theory to include viscous dissipation was first proposed by Kuwabara \& Kono \cite{Kuwabara87} (see also \cite{Brilliantov96,Morgado97}). It is assumed that the total force exerted on the objects can be decomposed into an elastic part and a dissipative part. Let $\delta(t)$ be the compression of the spheres at any instant $t$ and $\delta'=\frac{d}{dt}\delta(t)$, then the elastic contribution is Hertzian and proportional to $\delta^{3/2}$, and the dissipative contribution, which takes the form of the time derivative of the elastic part, is proportional to $\delta^{1/2}\delta'$. In terms of the dimensionless length $x=\delta/\delta_e$ and time $\tau=tV/\delta_e$, Newton's second law of motion is:
\begin{equation}\label{goveqn2}
\begin{split}
	\ddot x(\tau)+K&x^{1/2}(\tau)\dot x(\tau)+\frac54x^{3/2}(\tau)=0, \\
	& x(0)=0,\quad  \dot x(0)=1,
\end{split}
\end{equation}
where the over-dot denotes $d/d\tau$. The dimensionless quantity $K$ is given by
\begin{equation}\label{keypar1}
	K
	= \alpha\left(\frac{\eta^5VR}{M^2E^3}\right)^{1/5}, \quad \text{with}\ \ \alpha\approx1.924,
\end{equation}
where $\eta$ is the effective viscosity determined by the volume and shear coefficients of viscosities of the objects \cite{Brilliantov96}.



The parameter $K$ provides the ratio of viscous energy dissipation to the initial kinetic energy, as can be seen as follows: the viscous energy dissipation $U_1$ scales as $\eta T\int_\Omega|\nabla\mathbf u|^2 \text{d}x$ and the initial kinetic energy $U_2$ as $MV^2$, where $T$ is the collision time, $\Omega$ the volume over which dissipation occurs, and $\mathbf u$ the velocity field. For a central collision, the deformation of the spheres is concentrated in a volume $\Omega\sim a^3$ with $a$ the contact radius \cite{Leroy84}, and thus $|\nabla \mathbf u|\sim V/a$. This leads to $U_1/U_2\sim K.$ Since $K\propto 1/R$ for a given material, larger bodies dissipate less energy (notice that $M\propto R^3$).


In the absence of dissipation, $K=0$, and solving \eqref{goveqn2} leads to Hertzian dynamics as embodied in \eqref{basqua1}. When $K\ll 1$, corresponding to small viscous energy dissipation, we adopt the asymptotic solutions of \eqref{goveqn2} developed in \cite{Schwager98} and \cite{Schwager08}, summarized as follows. The trajectory $x(\tau)$ is written as a power series: $x(\tau)=\sum_{n=0} c_n \tau^{n/2}$, where the coefficients $c_n$ can be solved order by order. The reason $x(\tau)$ is expanded in $\tau^{1/2}$ instead of $\tau$ is that the derivatives higher than or equal to $x'''(\tau)$ diverge at $\tau=0$, as revealed by a closer examination of \eqref{goveqn2}. Once the trajectory $x(\tau)$ is known, the collision time $T$, maximum contact area $A$, and $\varepsilon$ (COR) can be calculated as 
\begin{align}
	T &=T_e[1+0.1009K+0.0576K^2+O(K^3)], \label{tasy1}\\
	A &= A_e[1-0.4036K+0.1817K^2+O(K^3)], \\
	\varepsilon &=1-1.0089K+0.6107K^2+O(K^3). \label{corasy1}
\end{align}
These expressions have been obtained in the literature in different forms (for example, see refs \cite{Brilliantov96,Schwager98,Schwager08}). Note that since $K\propto V^{1/5}$, $\varepsilon$ has a velocity dependence. The higher orders of $T$, $A$ and $\varepsilon$ can be obtained in a systematic way \cite{Schwager08}. The asymptotic series \eqref{tasy1}-\eqref{corasy1} appear to converge slowly, but for small values of $K$ our numerical solutions show that they give satisfactory approximations. An inspection of the coefficients in \eqref{tasy1}-\eqref{corasy1} shows that $T$ and $A$ (also the maximum compression, which is not shown here) depend less strongly on $K$ than $\varepsilon$ does. This is why the applicability of Hertz's theory extends into the inelastic regime.

\begin{figure}[h!]
\begin{center}
\includegraphics[width=3in]{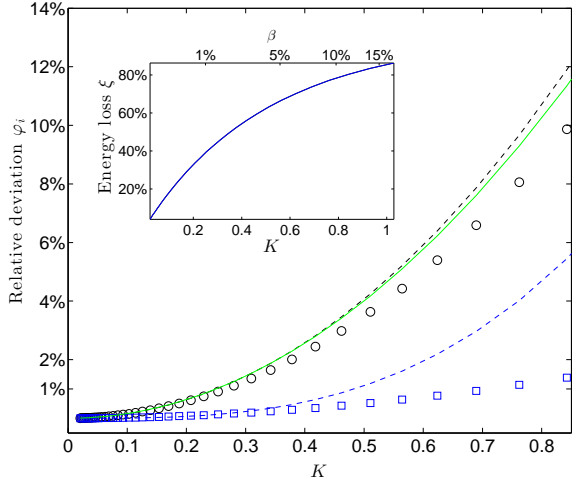}
\caption{Plots of $\varphi_1$ (black dashed) and $\varphi_2$ (blue dashed) as functions of $K$ from \eqref{2quant1}. Symbols are from the numerical solutions to \eqref{goveqn2}, which are bounded above by the green curve $K=2.5\beta^{1/2}$. 
Inset: Viscous energy loss $\xi$ as a function of $K$ and the allowable error $\beta$.}
\label{figerrK1}
\end{center}
\end{figure}

When comparing the theory with the experiments, it is more convenient to use the velocity-independent ratios  $Q_1\equiv A^{1/2}T^2$ and $Q_2\equiv A/\sqrt{VV^{\prime}}T$, where $V^{\prime}=\varepsilon V$ is the recoil velocity. When $K=0$ (i.e., elastic collisions), $Q_1=Q_{1e}\equiv A_e^{1/2}T_e^2=14.39M/E$, and $Q_2=Q_{2e}\equiv A_e/VT_e=1.0674 R$. When $K\neq 0$, we can use \eqref{tasy1}-\eqref{corasy1} to obtain
\begin{subequations}\label{2quant1}
\begin{equation}\label{q1}
	Q_1= Q_{1e}[1+0.1551K^2+O(K^3)],
\end{equation}
\begin{equation}\label{q2}
	Q_2= Q_{2e}[1-0.0031K^2+O(K^3)].
\end{equation}
\end{subequations}
We see that the linear terms in \eqref{2quant1} vanish and the coefficients of the quadratic and cubic terms are small, therefore $Q_i$ deviates only slightly from $Q_{ie}$ even if $K$ is relatively large. To see this more quantitatively, we define the relative deviation $\varphi_i \equiv |Q_i-Q_{ie}|/|Q_{ie}|$, $(i=1,2),$ and plot them in Fig.~\ref{figerrK1}. We can see that $\varphi_1$ and $\varphi_2$ are less than $1\%$ when $K\lesssim 0.25$, corresponding to $\varepsilon\gtrsim 0.78$ and fractional energy loss $\xi\equiv 1-\varepsilon^2\lesssim 39\%$ (Fig.~\ref{figerrK1} inset). Therefore using Hertz's theory results in no more than 1\% error even though nearly 40\% of the initial energy is lost, which is consistent with the experiments of Gugan \cite{Gugan00}. The physical basis for the robustness of $Q_i$ is that energy loss tends to decrease $A$ and $V^{\prime}$, but increase $T$. These deviations fortuitously offset each other to give a quadratic correction in $Q_i$. This result holds regardless of the size and material properties of the colliding bodies.

\section{Determining critical elastic velocity}


As $V$ increases above a certain value, the energy loss is so large that Hertz's theory is no longer a good approximation. We define $V_c(\beta)$ as the \emph{critical elastic velocity} below which $\varphi_1$ and $\varphi_2$ are smaller than an allowable error $\beta$, which in practice can be given by the precision of the measurements; from Fig.~\ref{figerrK1}, this corresponds to $K\lesssim 2.5\beta^{1/2}$. Using \eqref{keypar1} we can write
\begin{equation}\label{maxelavel1}
	V_c = 3.7\beta^{5/2}\Vd,
\end{equation}
where $\Vd = M^2E^3/R\eta^5$ is a characteristic velocity determined by the viscosity $\eta$.

\begin{figure}[h!]
\begin{center}
\includegraphics[width=3in]{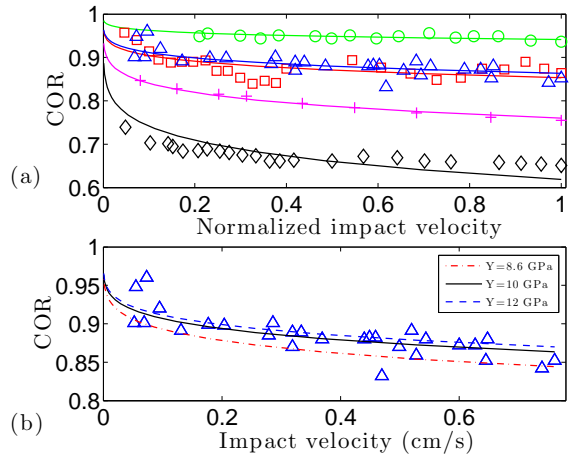}
\caption{(a) Fitting of the experimental data with the asymptotic expression \eqref{corasy1} using a least squares method. The impacting materials are steel ($\circ$), brass ($\square$), cork ($\Diamond$), ice with smooth surface ($\triangle$) and Jaques ball (+). For each data set, the impact velocity is normalized by the maximum experimental velocity which is not shown here. The values of $\eta$ giving the best fit are listed in Table~\ref{tab1}. (b) Most data points for smooth-surface ice collisions are constrainted by allowing $Y$ to vary between $Y_l=8.6$\,GPa and $Y_u=12$\,GPa (see text).}
\label{figcomdat1}
\end{center}
\end{figure}

In order to determine $\eta$ we fit the COR given by \eqref{corasy1} to experimental data. The results are given in Table~\ref{tab1}. The data in collisional experiments are somewhat scattered, but when $V$ is not too high we find consistently good agreement. We show in Fig.~\ref{figcomdat1}(a) the least squares fit of $\eta$, which provides 95\% confidence limits assuming a normal distribution. Then $V_c(\beta)$ can be calculated and that for $\beta=1\%$ and $5\%$ are given in Table~\ref{tab1}. Note that these values are obtained without considering the material yield strength. We can estimate, for example, that two steel spheres impacting at $V=V_c(1\%)$ experience a maximum compressive stress of about 3\,GPa, which is much greater than the yield strength of steel ($\sim 10^2$\,MPa). Other materials are likely to undergo plastic deformation at such high impact velocity as well. This implies that Hertz's theory is expected to be accurate until plastic deformation occurs.

\begin{table}[h!]
\begin{center}
\resizebox{3in}{!} {
  \begin{tabular}{| c | c | c | c | c | c |}
    \hline
	 & Steel & Brass & Cork & Ice & Jaques ball$^*$ \\ \hline 
	$E$ (GPa)  & 109 & 62 & 0.025 & 5.5 & 0.13 \\ [1.5ex] \hline
	$R$ (cm)  & 0.635 & 0.75 & 0.83 & 2.6 & 4.6 \\ [1.5ex] \hline
         $\eta\,(\times 10^5$ poise) & 6.7 & 15.4 & 0.25 & 9.62 & 1.47 \\ [1.5ex] \hline
	$V_c(1\%)$ (cm/s) & 2.4$\times 10^4$ & $3.5\times 10^3$ & 15.4 & 26.9 & $1.8\times10^3$\\ [1.5ex] \hline
	$V_c(5\%)$ (cm/s) & 1.4$\times 10^6$ & $2.0\times 10^5$ & 859 & $1.5\times 10^3$ & $9.8\times10^4$\\ [1.5ex] \hline
  \end{tabular}
  }
\end{center}
\caption{Estimated values of $\eta$, and $V_c(\beta)$ for $\beta=1\%$ and $5\%$ for different materials: steel \cite{Lifshitz64}, brass and cork \cite{Kuwabara87}, ice \cite{Bridges96} and Jaques ball \cite{Gugan00}. Note $(*)$: in \cite{Gugan00} a Jaques ball was impacting a steel plate, so the viscous dissipation in the steel is neglected compared to that in the Jaques ball.}\label{tab1}
\end{table}

\begin{figure}[h!]
\begin{center}
\includegraphics[width=3in]{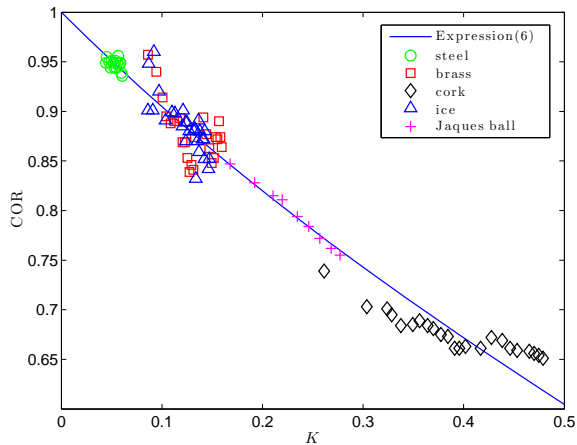}
\caption{Collapse of the data from Fig.~\ref{figcomdat1}(a), indicating that \eqref{corasy1} is valid for most impacting materials examined here (except for cork, in which the energy loss is too large and a systematic deviation exists).}
\label{figcollapCOR}
\end{center}
\end{figure}

The good agreement shown in Fig.~\ref{figcomdat1}(a) indicates that viscous dissipation is the dominant mechanism of energy loss in these experiments, and eq.~\eqref{corasy1} is a valid expression for the collision dynamics (except for cork, where maximum value of $K$ is about 0.49 and energy loss is nearly 60\%). Indeed, using the values of $\eta$ in Table~\ref{tab1} we can collapse these data onto a single curve, where the rescaled impact velocity $1.924(V/\Vd)^{1/5}=K$ is used (Fig.~\ref{figcollapCOR}).

The COR data for ice, despite its relatively large scatter, can also be well fitted by our model. It is known that polycrystalline ice has a range of measured elastic modulus (8.6\,GPa$\leq Y\leq$12\,GPa), depending on the grain size and shape, and the direction of the strain \cite{Schulson99}. Using $\eta=9.62\times 10^5$\,poise from Table~\ref{tab1} we are able to constrain most of the data by the lower (8.6\,GPa) and upper bound (12\,GPa) of the modulus (see Fig.~\ref{figcomdat1}b). Near $V=0$ the errors are larger due to greater uncertainties in measuring very small impact and recoil velocities, as well as contributions from surface frost \cite{Bridges96} and surface irregularities \cite{Chokshi93}.

\section{Evolution of energy loss}

\begin{figure}[h!]
\begin{center}
\includegraphics[width=2.8in]{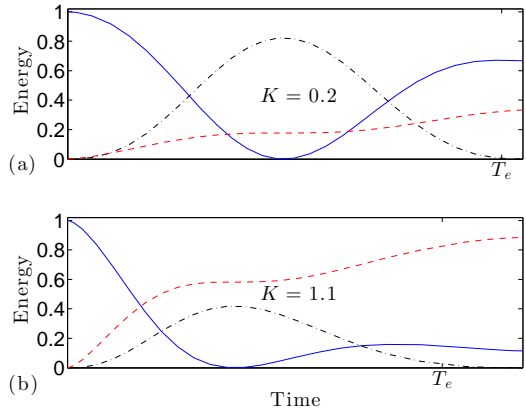}
\caption{The time evolution of kinetic energy $\Delta_k$ (solid), elastic energy $\Delta_e$ (dot-dash) and dissipated energy $\Delta_d$ (dashed), for small (a) and large (b) energy dissipation. The maximum compression and the end of collision correspond to when $\Delta_k$ and $\Delta_e$ become zero, respectively.}
\label{figenloss}
\end{center}
\end{figure}

Let $\Delta_k$, $\Delta_e$ and $\Delta_d$ be respectively the kinetic, elastic, and dissipated energy of the system, normalized by the initial kinetic energy. Clearly $\Delta_k+\Delta_e+\Delta_d=1$, and their time evolutions are plotted in Fig.~\ref{figenloss}. In the compression stage, $\Delta_k$ decreases with time and becomes zero when $\Delta_e$ reaches its maximum. After that, $\Delta_e$ decreases, converting the stored elastic energy into kinetic and dissipated energy. The dissipated energy $\Delta_d$ is always increasing. We find that near the end of recoil the objects may slow down slightly ($\Delta_k$ decreases), which means that the dissipated energy becomes so large. For small dissipation, both $\Delta_k$ and $\Delta_e$ are nearly symmetric, and the energy dissipation before and after the maximum compression are almost the same (Fig.~\ref{figenloss}a). However, for large dissipation $\Delta_k$ and $\Delta_e$ are asymmetric, and the majority of the energy dissipation takes place before the maximum compression is reached (Fig.~\ref{figenloss}b). Such an asymmetry has also been observed for soft-body collisions \cite{Tanaka05}. Finally, the collision time $T$ increases as $K$ increases, consistent with eq.~\eqref{tasy1}.

\section{Size dependence of fracturing}


When $V$ is sufficiently large, plastic deformation will be triggered, which take the form of irreversible dislocations for ductile materials and of crack formation for brittle materials. Since the creation of dislocations and cracks consumes a great amount of energy, the COR will drop abruptly when the ``fracturing velocity'' $V_F$ is reached \cite{Lifshitz64,Higa98}. While fracturing at a finite velocity is ubiquitous for all materials, it has been observed that larger ice spheres fracture at a lower $V_F$, as if they are weaker than the smaller ones \cite{Higa98}. Despite several explanations such as defect weakening and strain-rate hardening mentioned before, we provide an alternative explanation based on the viscoelastic model. The key fact is that a larger sphere loses less energy during the impact, and thus experiences a higher compressive stress at its maximum compression. 
Therefore a perfectly isotropic and homogeneous sphere in the absence of initial defects would still exhibit such a size dependence of $V_F$. 

\begin{figure}[h!]
\begin{center}
\includegraphics[width=2.5in]{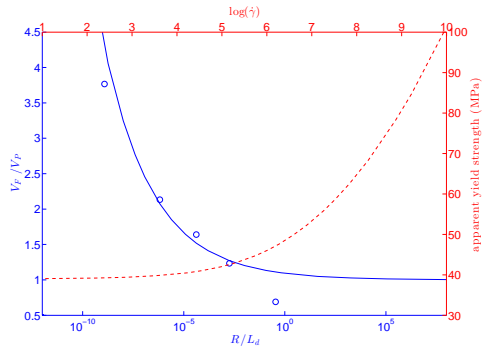}
\caption{(Blue axes) The dimensionless fracturing velocity decreases with the object size, and approaches 1 as $R/L_d$ approaches infinity. The circles are experimental data from Higa\cite{Higa98} and the blue line is the numerical solution. (Red axes) The dashed line represents the apparent yield strength of ice \emph{assuming no viscous dissipation}, which exhibits a strain-rate hardening phenomenon.}
\label{figfracv}
\end{center}
\end{figure}

Fracturing of a material emerges when the compressive stress exceeds the material yield strength $\tau^\star$. Since brittle materials, including ice and metals at low temperatures, undergo little or no plastic deformation before fracturing, it is reasonable to assume that $\tau^\star$ is equal to the maximum compressive stress at $V=V_F$. It is known that the maximum compressive stress is 0.31 of the maximum pressure $P_0$ for ice \cite{Johnson85}, so $\tau^\star$ is related to the $P_0$ through $\tau^\star=0.31P_0$. Therefore, for a given $\tau^\star$, $\Vf$ can be determined from the numerical solutions to \eqref{goveqn2}. 
Fig.~\ref{figfracv} (left and lower axes) shows the normalized fracturing velocity $\Vf/\Vp$ as a function of $R/L_d$, where $\Vp\equiv\left(\pi^5 R^3 P_0^5/30ME^4\right)^{1/2}$ is the size-independent Hertzian velocity that would give a maximum pressure $P_0$ in an elastic collision, and $L_d\equiv\eta/\rho\Vd$ is the viscous length. Since $\Vd\propto R^5$, $L_d$ has a strong size dependence, seen as the wide range on the abscissa of Fig.~\ref{figfracv}. The size dependence of the energy loss $\xi$ is reflected in $V_F$: as $R$ increases, $\xi$ decreases and thus $V_F$ approaches its elastic value $V_P$. The solid curve in Fig.~\ref{figfracv} represents $\tau^\star=40.3$\,MPa (or $P_0=130$\,MPa) and shows good agreement with the experimental data of Higa \emph{et al.} \cite{Higa98}. The largest deviation is found for $R=3.6$\,cm, where the ice spheres may be sufficiently large that defect weakening becomes important.


If viscous energy dissipation is neglected, Hertz's theory would predict an \emph{apparent} strain-rate hardening \cite{Jones07,Kim07}. Note that here the strain rate is referred to its maximum value as it is a function of time. The maximum strain rate is achieved at the maximum compression, which can be estimated as $\dot\gamma=0.69P_0/ET_e\propto V^{3/5}/R.$ Therefore smaller objects have a larger strain rate at a given impact velocity. Using the same parameters for ice as in \cite{Higa98}, we plot in Fig.~\ref{figfracv} (right and upper axes) the apparent yield strength \emph{as if}  viscous dissipation is absent. The increasing trend of the yield strength conforms to the strain-rate hardening observed in the Split-Hopkinson pressure bar (SHPB) test, and the values of the apparent yield strength (several tens of MPa) are in good agreement with the maximum compressive strength of ice in the regime of high strain rate \cite{Kim07}.

%

\section{Conclusion}


Using a viscoelastic model we have explained why Hertz's elastic theory is valid far beyond its elastic limit in modeling collisional dynamics. In situations that viscous dissipation is the dominant mechanism of energy loss, the collision time and maximum contact area have a weak dependence on energy loss, with deviations of 3\% and 10\%, respectively, when the energy loss is nearly 40\%. Moreover, the two velocity-independent ratios $Q_{1}$ and $Q_{2}$ have even weaker dependence on the energy loss, with deviations less than 1\% for the same energy loss. By fitting to the experiments in the literature, we have estimated the effective viscosity for different materials and the critical impact velocity below which Hertz's theory is practically useful. For most materials we examined, Hertz's theory gives less than a few percent error in $Q_1$ and $Q_2$ even when the materials fail.

The key parameter $K$ characterizing the fractional energy loss increases with the impact velocity and viscosity, and decreases with effective Young's modulus. Since $K$ is also inversely proportional to $R$, larger colliding objects lose less energy and thus have a higher COR than smaller ones. This trend has been observed experimentally \cite{Labous97}, and is used by us to explain the size dependence of the critical velocity $V_F$ at which fracturing occurs. It is noteworthy, however, that this size dependence is too strong and does not fit some of the COR data very well. This is likely because when deriving eq.~\eqref{goveqn2} the elastic part of the force is assumed to be Hertzian even though the deformation is very large. In reality, the force-displacement curve should grow more slowly than $\delta^{3/2}$, and hence very small particles do not dissipate so much kinetic energy. 


Perhaps even further beyond our expectations, Hertz's theory seems to extend its validity to fluid mechanics. Molotskii \emph{et al.} found that for a liquid droplet with radius $R$ sitting on a superhydrophobic surface the contact area $A$ scales as $A\sim R^{10/3}$ \cite{Molotskii09}. If $\sigma_l/R$ is used as the effective Young's modulus where $\sigma_l$ is the liquid-air surface tension, the data fit Hertzian theoretical predictions very well. The power $R^{10/3}$ lies between $R^3$ and $R^4$, which are the scalings of the contact area for droplets that are much smaller and larger than the capillary length, respectively \cite{Mahadevan99,Taylor73}. Since the droplets used by Molotskii \emph{et al.} \cite{Molotskii09} are comparable to the capillary length, Hertz's theory seems to be valid for an intermediate regime where the effects of surface tension and gravity are comparable. Whether or not this implies a deeper connection between Hertz's theory and fluid mechanics, a classical theory once again proves its lasting merit in inspiring us to better understand the nature beyond our current belief.

\section{Acknowledgments}
We thank Rod Cross and Thorsten P\"oschel for enlightening discussions. The authors acknowledge the financial support from the Swedish Research Council and Yale University and additionally JSW thanks the Wenner-Gren Foundation, the John Simon Guggenheim Foundation, and a Royal Society Wolfson Research Merit Award.

\bibliography{rbib}



\end{document}